# ESO observing programme FDS (Fornax Deep Survey)

## Abstract

We present the Fornax Deep Survey (FDS), an imaging survey using using the wide-field imager OmegaCAM mounted on the VST in the SDSS u', g', r', and i'-bands of about 30 square degrees covering the Fornax Galaxy Cluster and the infalling Fornax A Group. FDS is a joint project that used the VST-GTO time of NOVA (previously called FOCUS - PI: R. F. Peletier) and INAF (as part of VEGAS - PIs: M. Capaccioli and E. Iodice). This survey makes it possible to study the galaxies down to r'-band magnitude $m_{r'} \approx 21$ mag ($M_{r'} \approx -10.5$ mag), which opens a new parameter regime to investigate the evolution of dwarf galaxies in the cluster environment. After the Virgo cluster, Fornax is the second nearest galaxy cluster to us, and with its different mass and evolutionary state, it provides a valuable comparison that makes it possible to understand the various evolutionary effects on galaxies and galaxy clusters. These data provide an important legacy dataset to study the Fornax cluster. Details about the survey can be found in A. Venhola, R. F. Peletier, E. Laurikainen et al., 2018, A&A 620, 165. Here the imaging fits-files are presented. In data release 2, a catalog containing a complete sample of dwarf galaxies in Fornax will be added (from A. Venhola, R. F. Peletier, E. Laurikainen et al., 2019, A&A, 625, A143), as well as a catalog of sources, such as globular clusters in the Fornax Cluster and background galaxies, (from M. Cantiello, A. Venhola, A. Grado et al., 2020, A&A, 639, A136).

# Overview of Observations

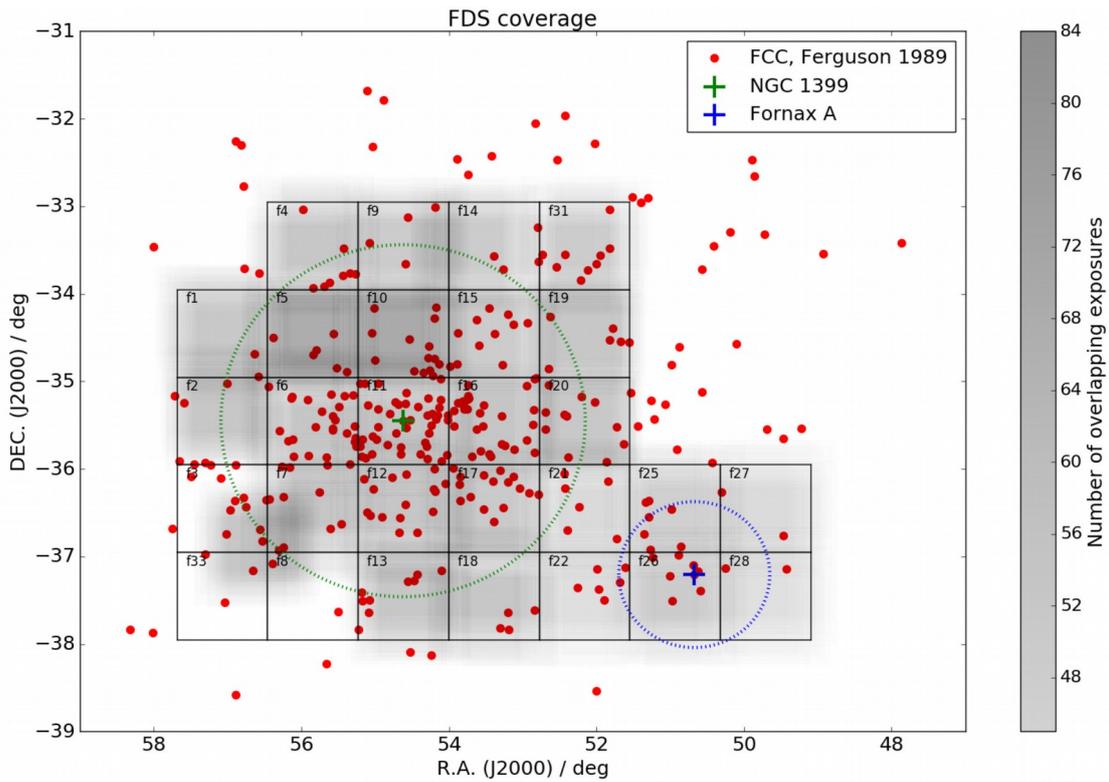

**Figure 1:** Map of the Fornax Cluster indicating the various tiles of 1°x1° (corresponding to about 325 kpc x 325 kpc at the distance of the Fornax cluster). The r′-band weight maps are shown in gray-scale: darker color corresponding to deeper observations. All the FCC galaxies (Ferguson, 1989, AJ 98, 367) classified as "likely members" or "definitive members" are shown with red points. The green dotted circle shows the virial radius of 2.2° (≈ 0.7 Mpc), and the green cross shows the central galaxy NGC 1399. The blue cross and the blue-dotted line show the peculiar elliptical galaxy NGC 1316 at the center of the Fornax A subgroup, and the 2σ galaxy overdensity around it, respectively.

The cluster has been observed in u', g', r' and i'. The surveyed area is indicated in Figure 1, including the tiles in which it has been organised.

# Release Content

The depth of the individual tiles, as well as their seeing, are indicated in Table 1 (from Venhola et al. 2018). In this table the first column gives the name of the field, the four next columns give the mean and the RMS of the FWHM within the field in the different bands, and the four last columns show the surface brightness corresponding to 1σ S/N per pixel for a given field in the different photometric bands. One pixel corresponds to 0.20''. One can see in this table that there are no data for Field 8, no u'-band observations for the Fornax A region (Fields 22, 25-28), and that there are only g' and r'-band observations for Field 33, an outlying field. Note that in the corner of Field 33, a small area has not been observed because of the presence of a very bright star.

| field | FWHM ± $\sigma_{FWHM}$ / arcsec | | | | depth / mag arcsec$^{-2}$ | | | |
|---|---|---|---|---|---|---|---|---|
| | u' | g' | r' | i' | u' | g' | r' | i' |
| Field1 | 1.17±0.09 | 1.35±0.09 | 1.14±0.15 | 0.69±0.07 | 25.16 | 26.76 | 26.05 | 25.24 |
| Field2 | 1.21±0.04 | 1.11±0.07 | 0.90±0.07 | 0.79±0.08 | 25.14 | 26.60 | 26.03 | 25.00 |
| Field4 | 1.18±0.11 | 1.39±0.05 | 1.19±0.12 | 0.70±0.07 | 25.20 | 26.70 | 26.01 | 25.26 |
| Field5 | 1.33±0.07 | 1.15±0.10 | 1.39±0.14 | 1.08±0.15 | 25.58 | 26.79 | 26.10 | 25.22 |
| Field6 | 1.11±0.05 | 0.84±0.08 | 1.08±0.08 | 1.21±0.12 | 25.68 | 26.72 | 25.98 | 25.07 |
| Field7 | 1.04±0.05 | 0.83±0.10 | 0.95±0.09 | 1.42±0.11 | 25.55 | 26.81 | 26.06 | 24.87 |
| Field9 | 1.38±0.07 | 1.20±0.08 | 0.97±0.09 | 0.84±0.08 | 25.30 | 26.83 | 26.15 | 25.37 |
| Field10 | 1.34±0.05 | 1.15±0.04 | 1.02±0.12 | 1.09±0.07 | 25.66 | 26.77 | 26.16 | 25.24 |
| Field11 | 1.27±0.05 | 1.06±0.12 | 1.09±0.11 | 1.15±0.06 | 25.25 | 26.51 | 26.02 | 25.04 |
| Field12 | 1.15±0.06 | 0.83±0.10 | 1.04±0.10 | 1.17±0.10 | 25.69 | 26.74 | 26.09 | 25.04 |
| Field13 | 1.10±0.05 | 0.91±0.06 | 1.03±0.06 | 1.16±0.07 | 25.39 | 26.83 | 26.14 | 25.44 |
| Field14 | 1.34±0.06 | 1.18±0.08 | 0.96±0.09 | 0.86±0.07 | 25.28 | 26.70 | 26.00 | 25.29 |
| Field15 | 1.30±0.04 | 1.13±0.05 | 0.90±0.07 | 0.97±0.06 | 25.37 | 26.60 | 26.14 | 25.12 |
| Field16 | 1.31±0.04 | 1.26±0.07 | 0.94±0.08 | 1.08±0.09 | 25.52 | 26.68 | 26.09 | 25.21 |
| Field17 | 1.27±0.04 | 1.11±0.12 | 0.87±0.08 | 1.01±0.08 | 25.35 | 26.54 | 26.21 | 25.17 |
| Field18 | 1.11±0.06 | 0.95±0.08 | 1.03±0.09 | 1.12±0.11 | 25.33 | 26.79 | 26.17 | 25.43 |
| Field19 | 1.26±0.04 | 1.14±0.13 | 0.89±0.07 | 0.87±0.08 | 25.25 | 26.70 | 26.14 | 25.23 |
| Field20 | 1.30±0.06 | 1.22±0.07 | 0.95±0.09 | 1.08±0.07 | 25.29 | 26.46 | 26.06 | 25.04 |
| Field21 | 1.22±0.05 | 1.12±0.06 | 0.78±0.05 | 0.88±0.07 | 25.13 | 26.51 | 25.84 | 25.28 |
| Field22 | -±- | 1.04±0.07 | 0.81±0.05 | 0.85±0.07 | - | 26.52 | 25.90 | 25.16 |
| Field25 | -±- | 1.11±0.10 | 0.77±0.06 | 0.85±0.07 | - | 26.63 | 25.84 | 25.11 |
| Field26 | -±- | 0.93±0.07 | 0.81±0.05 | 0.91±0.07 | - | 25.89 | 25.96 | 25.06 |
| Field27 | -±- | 1.07±0.10 | 0.78±0.06 | 0.89±0.10 | - | 26.39 | 25.63 | 24.88 |
| Field28 | -±- | 1.08±0.14 | 0.79±0.09 | 0.92±0.09 | - | 26.31 | 25.57 | 24.89 |
| Field31 | 1.33±0.05 | 1.22±0.13 | 1.00±0.08 | 0.86±0.08 | 25.11 | 26.58 | 25.86 | 24.98 |
| Field33 | -±- | 1.09±0.07 | 0.84±0.07 | -±- | - | 26.40 | 25.74 | - |

**Table 1:** effective seeing and depth of the various fields in the 4 bands (from Venhola et al. 2018).

The survey release consists of 97 science files and 97 weight files. The total (compressed) data volume is 182 GB.

# Release Notes

## Data Reduction and Calibration

The data reduction and calibration is explained in detail in Venhola et al. (2018). Here a summary is given.

The **observing strategy** of the FDS is described in Venhola et al. (2017, A&A, 608, A142) and Iodice et al. (2016, ApJ, 820, 42), but for completeness, a short description is given also here. The observations were performed using short exposure times of approximately 3 minutes, and large (≈1 deg) dithers between the consecutive exposures. The fields were observed in sets of two to three fields in such a way that after visiting all the fields once, an offset of ~10 arcmin with respect to the previous observation of a given field was made. The directions of these small offsets were randomly chosen around the centers of the fields. The large dithers and offsets ensure that the same objects do not appear twice in the same pixel, and makes it possible to stack consecutive observations as a background model.

For reference, the halo of NGC 1399, located in Field 11, extends over an area of 1 deg² (Iodice et al. 2016), which would lead to it covering the full field of view of the observations of that field if we did not use the adopted dithering and offset strategy. To obtain the necessary depth in the images each field was visited 75, 55, 55, and 35 times with the u′,g′,r′, and i′ filters, respectively. The observations cover a 20 deg² area in the main cluster in u′,g′,r′, and i′, and an additional 6 deg² in the Fornax A South West subgroup in g′,r′, and i′. All observations follow a regular grid of target fields comprising continuous coverage, except in the area of Fields 3 and 33, in which some gaps occur due to bright stars.

The data was reduced using the AstroWISE system, as was done, for example, for the KIDS survey. The **instrumental corrections** applied for each frame include overscan correction, removal of bias, flatfielding, illumination correction, masking of the bad pixels, and subtraction of the background. The data is overscan corrected by subtracting from each pixel row the row-wise median values, read from the CCD overscan areas. The fine structure of the bias is then subtracted using a master bias frame stacked from ten overscan corrected bias frames. Flatfielding is done after bias correction using a master flat-field which is combined from eight twilight flatfields and eight dome flatfields. Before combining the different flatfields, the high spatial frequencies are filtered out from the twilight flat-fields, and the low frequency spatial Fourier frequencies from the dome flatfields. This approach is adopted, since the dome flatfields have better signal-to-noise ratios to correct for the pixel-to-pixel sensitivity variations, whereas the twilight flat-fields have more similar overall illumination with the science observations. During the instrumental reduction, weight maps are also created for each individual frame. Weight maps carry information about the defects or contaminated pixels in the images and also the expected noise associated with each pixel. The hot and cold pixels are detected from the bias and flatfield images, respectively. These pixels are then set to zero in the weight maps. The flatfielded and debiased images are also searched for satellite tracks and cosmic rays, and the values of the pixels in the weight maps corresponding to the contaminated pixels in the science images, are then set to zero. The Hough transformation method (Vandame 2001, Mining the Sky, p. 595) is applied to the images to pick up the satellite tracks, which are eliminated by masking the lines consisting of more than 1000 pixels that have intensity above the 5-$\sigma$ level relative to the background and are located on the same line. Cosmic rays are detected using SExtractor, and the corresponding pixels are masked from the weight maps. The pixels in the weight maps have values $W = M_{bad} / \sigma^2$, where $\sigma$ is the standard deviation of the background noise and $M_{bad}$ is the combined bad pixel map where the bad pixels have been set to zero and other pixels to one. The observations contain an additional smooth light component resulting from scattered light. A careful removal of this component is essential for studying the outskirts of the galaxies and the low surface brightness objects. A background model is created first by scaling a set of 12 consecutive exposures of the targets, and then median averaging the stack. The scaling factors between images A and B is defined by measuring median values within small boxes in image A ($m_A$), and in the same locations in image B ($m_B$), and then taking the median of their ratios: $s = \mathrm{median}(m_A/m_B)$ For each image among those to be stacked, such a scaling factor is defined with respect to A, and the images are multiplied withthese factors before stacking. If there is a large scatter between the ratios of *s*, the chip medians of the exposures are scaled with each other. The scaled images are then median stacked to the background model, and the model is subtracted from image A. This strategy allows us also to remove the fringe patterns appearing in the OmegaCAM *i'*-band images, and removes also all the possible residual patterns from theflatfielding. Systematic photometric residual patterns still remain after flatfielding, which are corrected by applying an illumination correction to the data. We used the correction models made for the Kilo Degree Survey (KiDS; see Verdoes Kleijn et al. 2013, Experimental Astronomy, 35, 103 for details). The models were made by mapping the photometric residuals across the OmegaCAM's CCD array using a set of dithered observations of Landolt's Selected Area (SA) standard star fields (A.U. Landolt, 1992, AJ, 104, 340), and fitting a linear model to the residuals. The images were multiplied with this illumination correction. The illumination correction is applied after the background removal to avoid producing artificial patterns into the background of images.

As far as the **astronometric calibration** is concerned: the reduced images are calibrated to world coordinates using SCAMP (Bertin 2006, ASP 351, p. 112). We make the coordinate transformation by applying first the shifts and rotations according to the image headers. The fine tuning of the astrometric calibration is obtained by first associating the source lists extracted from the science images with the 2 Micron All-Sky Survey Point Source Catalog (2MASS PSC; Cutri et al. 2003) and fitting the residuals by a second order polynomial plane. This polynomial correction is then applied to the data coordinates, and the pixel size is sampled to 0.20 arcsec pixel$^{-1}$. After applying the astrometric calibration, the remaining differences between the 2MASS PSC objects and the corresponding objects in our data have root mean square (RMS) of 0.1 arcsec.

The absolute **zero point calibration** is done by observing standard star fields each night and comparing their OmegaCAM magnitudes with the Sloan Digital Sky Survey Data Release 11 (SDSS DR11; Alam et al. 2015, ApJS, 219, 12) catalog values. The OmegaCAM point source magnitudes are first corrected for the atmospheric extinction by subtracting a term *kX*, where *X* is airmass and *k* is the atmospheric extinction coefficient with the values of 0.515, 0.182, 0.102 and 0.046 for u′, g′, r′ and i′, respectively. The zero-point for a given CCD is the difference between the corrected magnitude of the object measured from a standard starfield exposure and the catalog value. The zero-points are defined only once per night, so that for each science observation only the varying airmass was corrected.

The calibrated exposures are median stacked into **mosaic images** using SWarp (Bertin 2010), and the contaminated pixels are removed using the weight maps. SWarp produces also a mosaic weight map for each mosaic, where the pixel values are inverse of the variance associated to each pixel. We stack the images according to the FDS fields with an extra overlap of 5 arcmin on each side, so that we do not need to cut any large galaxies later in the analysis. As a final result we produce 1.17°×1.17° mosaics and the corresponding weight images.

## Data Quality

To measure the image **depth** we used the final mosaics to measure the actual obtained depth in the images. To measure the background noise in the images, we defined 500 boxes with 200×200 pixels in size, randomly distributed in the images, and calculate the three times σ-clipped standard deviations of the pixel values within the boxes. As the final σ-value of each field we take the median of the calculated standard deviations. The measured σs for all fields in the different bands are listed in Table 1. We find that the obtained depth in the images for 1σ signal-to-noise per pixel corresponds to median surface brightness values of 26.6, 26.7, 26.1, and 25.5 mag arcsec$^{-2}$ in u′, g′, r′, and i′-bands, respectively. When averaged over an 1 arcsec$^2$ area, these values correspond to surface brightness of 28.3, 28.4, 27.8, 27.2 mag arcsec$^{-2}$ in u′, g′, r′, and i′, respectively.

As the Fornax cluster is poorly covered with standard star catalogs, a straightforward comparison of the obtained magnitudes with the standard stars to define the **photometric accuracy** is not possible. However, we can do an internal photometric consistency check by using the fact that the Milky-Way stars form locii in the color-color space that have constant locations and small intrinsic scatter. Ivezić et al. (2004, AN, 325, 583) have performed analysis for the zero-point accuracy of the SDSS, using the stellar locus in various color-color diagrams. Applying this analysis, we show that the errors associated to our zero point definitions are roughly three time as large as for the SDSS images, corresponding to 0.03 mag in g′, r′, and i′-bands and 0.04 in u′-band. Details of this analysis are given in Venhola et al. (2018).

## Known issues

_

## Previous Releases

None

## Data Format

### Files Types

The files are in FITS format, with the relevant information in the header. They have been compressed using NASA's HEASARC's fpack routine (https://heasarc.gsfc.nasa.gov/fitsio/fpack/). Each science frame is accompanied by a weight frame. Files are named based on the field covered and the filter used for observations following the format:

FDS_F<FieldNumber>_OCAM_<FilterName>_sci.fits.fz for science images and
FDS_F<FieldNumber>_OCAM_<FilterName>_wei.fits for weight maps.

### Catalogue Columns

None

## Acknowledgements

The FDS data were produced by the FDS collaboration, Reynier Peletier (PI,1), Enrichetta Iodice (PI,2) Aku Venhola (3), Massimo Capaccioli (2), Michele Cantiello (4), Raffaele D'Abrusco (5), Jesús Falcón-Barroso (6), Aniello Grado (2), Michael Hilker (7), Luca Limatola (2), Steffen Mieske (8), Nicola Napolitano (2,9), Maurizio Paolillo (10,2), Marilena Spavone (2), Edwin Valentijn (1), Glenn van de Ven (11) and Gijs Verdoes Kleijn (1).

(1) Kapteyn Institute, University of Groningen, Landleven 12, 9747 AD Groningen, the Netherlands
(2) INAF - Astronomical Observatory of Capodimonte, Salita Moiariello 16, I80131, Naples, Italy
(3) Astronomy Research Unit, University of Oulu, Pentti Kaiteran katu 1, 90014 Oulu, Finland
(4) INAF Osservatorio Astr. di Teramo, via Maggini, I-64100, Teramo, Italy
(5) Center for Astrophysics | Harvard & Smithsonian, 60 Garden St., 02138 Cambridge MA, USA
(6) Instituto de Astrofisica de Canarias, C/ Via L'actea s/n, 38200 La Laguna, Spain
(7) European Southern Observatory, Karl-Schwarzschild-Strasse 2, D-85748 Garching bei München, Germany
(8) European Southern Observatory, Alonso de Cordova 3107, Vitacura, Santiago, Chile
(9) School of Physics and Astronomy, Sun Yat-sen University Zhuhai Campus, 2 Daxue Road, Tangjia, Zhuhai, Guangdong 519082, China
(10) University of Naples Federico II, C.U. Monte Sant'Angelo, Via Cinthia, 80126 Naples, Italy
(11) Department of Astrophysics, University of Vienna, Türkenschanzstrasse 17, 1180 Wien, Austria

Several papers have been based on the FDS, which is based on data products created from observations collected at the European Organisation for Astronomical Research in the Southern Hemisphere under ESO programme(s) 60.A-9038(A), 088.B-4012(A), 090.B-0414(A), 092.B-0744(D), 092.B-0623(A), 094.B-0496(A), 094.B-0512(A), 094.B-0512(B), 096.B-0501(B), 096.B-

0582(A), 098.B-0208(A), 098.B-0298(A), 098.B-0298(B), 0100.B-0148(A), 0100.B-0148(B) and 0100.B-0168(A). These consist of core-papers, mainly based on the imaging, and others that are based on the FDS. At the moment of first release of the FDS imaging, the core papers are:

Iodice, E., Capaccioli, M., Grado, A., et al., 2016, The Fornax Deep Survey with VST. I. The Extended and Diffuse Stellar Halo of NGC 1399 out to 192 kpc, ApJ, 820, 42

Iodice, E., Spavone, M., Capaccioli, M., et al., 2017, The Fornax Deep Survey with VST. II. Fornax A: A Two-phase Assembly Caught in the Act, ApJ, 839, 21

Venhola, A., Peletier, R., Laurikainen, E., et al., 2017, The Fornax Deep Survey with VST. III. Low surface brightness dwarfs and ultra diffuse galaxies in the center of the Fornax cluster, A&A, 608, A142

Venhola, A., Peletier, R., Laurikainen, E., et al., 2018, The Fornax Deep Survey with the VST. IV. A size and magnitude limited catalog of dwarf galaxies in the area of the Fornax cluster, A&A, 620, A165

Iodice, E., Spavone, M., Capaccioli, M., et al., 2019, The Fornax Deep Survey with the VST. V. Exploring the faintest regions of the bright early-type galaxies inside the virial radius, A&A, 623, A1

Venhola, A., Peletier, R., Laurikainen, E., et al., 2019, The Fornax Deep Survey (FDS) with VST. VI. Optical properties of the dwarf galaxies in the Fornax cluster, A&A, 625, A143

Raj, M. A., Iodice, E., Napolitano, N. R., et al., 2019, The Fornax Deep Survey with the VST. VII. Evolution and structure of late type galaxies inside the virial radius of the Fornax cluster, A&A, 628, A4

Spavone, M., Iodice, E., van de Ven, G., et al., 2020, The Fornax Deep Survey with VST. VIII. Connecting the accretion history with the cluster density, A&A, 639, A14

Cantiello, M., Venhola, A., Grado, A., et al., 2020, The Fornax Deep Survey with VST. IX. Catalog of sources in the FDS area with an example study for globular clusters and background galaxies, A&A, 639, A136

Raj, M. A., Iodice, E., Napolitano, N. R., et al. 2020, The Fornax Deep Survey with VST. X. On the assembly history of the bright galaxies and intra-group light in the Fornax A subgroup, arXiv:2006.11828

We would like to thank Thorsten Lisker for the valuable help and discussions during the whole FDS project. We acknowledge financial support from the European Union's Horizon 2020 research and innovation programunder the Marie Skłodowska-Curie grant agreement No. 721463 to the SUNDIAL ITN network. JFB acknowledges support through the RAVET project by the grant AYA2016-77237-C3-1-P from the Spanish Ministry of Science, Innovation and Universities

(MCIU) and through the IAC project TRACES, which is partially supported through the state budget and the regional budget of the Consejería de Economía, Industria, Comercio y Conocimiento of the Canary Islands Autonomous Community. GvdV acknowledges funding from the European Research Council (ERC) under the European Union's Horizon 2020 research and innovation programme under grant agreement No 724857 (Consolidator Grant ArcheoDyn). NRN, MP, and EI acknowledge support from PRIN INAF 2014 "Fornax Cluster Imaging and Spectroscopic Deep Survey".